\documentclass[runningheads]{llncs}
\usepackage{graphicx}
\usepackage{amsmath}
\usepackage{marvosym}
\usepackage{algorithm}
\usepackage{algorithmicx}
\usepackage{algpseudocode}
\usepackage{amsmath}
\usepackage{amssymb}
\usepackage{cite}
\usepackage{booktabs}
\usepackage{bm}
\usepackage{tabularx}
\usepackage{makecell}
\usepackage{hyperref}
\usepackage{fancyhdr}
\usepackage{color}
\usepackage{ulem}
\usepackage{url}
\usepackage{multirow}
\makeatletter
\newcommand{\printfnsymbol}[1]{%
  \textsuperscript{\@fnsymbol{#1}}%
}

\begin{document}

\title{CarveMix: A Simple Data Augmentation Method for Brain Lesion Segmentation}

\author{Xinru Zhang$^{\mathrm{1},}$\thanks{Equal contribution; \textsuperscript{\Letter} co-corresponding authors.}, Chenghao Liu$^{\mathrm{1},}$\printfnsymbol{1}, Ni Ou$^{\mathrm{2}}$, Xiangzhu Zeng$^{\mathrm{3}}$, Xiaoliang Xiong$^{\mathrm{4}}$, Yizhou Yu$^{\mathrm{4}}$, Zhiwen Liu$^{\mathrm{1}}$\textsuperscript{(\Letter)}, Chuyang Ye$^{\mathrm{1}}$\textsuperscript{(\Letter)}}


\institute{$^{1}$School of Information and Electronics, Beijing Institute of Technology, Beijing, China\\ 
\email{\{zwliu,chuyang.ye\}@bit.edu.cn}\\
$^{2}$School of Automation, Beijing Institute of Technology, Beijing, China\\
$^{3}$Department of Radiology, Peking University Third Hospital, Beijing, China\\
$^{4}$Deepwise AI Lab, Beijing, China
}

\authorrunning{Zhang and Liu et al.}

\titlerunning{CarveMix}
%
%
%
\maketitle             
\begin{abstract}
Brain lesion segmentation provides a valuable tool for clinical diagnosis, and \textit{convolutional neural networks}~(CNNs) have achieved unprecedented success in the task. 
Data augmentation is a widely used strategy that improves the training of CNNs, and the design of the augmentation method for brain lesion segmentation is still an open problem.
In this work, we propose a simple data augmentation approach, dubbed as CarveMix, for CNN-based brain lesion segmentation.
Like other ``mix"-based methods, such as Mixup and CutMix, CarveMix stochastically combines two existing labeled images to generate new labeled samples. Yet, unlike these augmentation strategies based on image combination, CarveMix is lesion-aware, where the combination is performed with an attention on the lesions and a proper annotation is created for the generated image.
Specifically, from one labeled image we carve a \textit{region of interest}~(ROI) according to the lesion location and geometry, and the size of the ROI is sampled from a probability distribution.
The carved ROI then replaces the corresponding voxels in a second labeled image, and the annotation of the second image is replaced accordingly as well.
In this way, we generate new labeled images for network training and the lesion information is preserved.
To evaluate the proposed method, experiments were performed on two brain lesion datasets. The results show that our method improves the segmentation accuracy compared with other simple data augmentation approaches. 

\keywords{Brain lesion segmentation \and Data augmentation \and Convolutional neural network.}
\end{abstract}
\section{Introduction}

Quantitative analysis of brain lesions may improve our understanding of brain diseases and treatment planning~\cite{IKRAM2010378,validity}.
Automated brain lesion segmentation is desired for reproducible and efficient analysis of brain lesions, and \textit{convolutional neural networks}~(CNNs) have achieved state-of-the-art performance of brain lesion segmentation~\cite{deepmedic,0nnU}.

Data augmentation is a widely used strategy for improving the training of CNNs, where additional training data is generated from existing training data. It is shown to reduce the variance of the mapping learned by CNNs~\cite{2019A} and has been effectively applied to brain lesion segmentation~\cite{2020Analyzing,0nnU}. 
Data augmentation can be achieved with basic image transformation, including translation, rotation, flipping, etc.~\cite{0nnU}, where an existing training image is transformed together with the annotation using hand-crafted rules. 
Since the diversity of the data generated via basic image transformation can be limited, more advanced approaches based on generative models have also been developed~\cite{2020Semi,adaptive}. However, the implementation of these methods is usually demanding, and the training of generative models is known to be challenging~\cite{0Embracing}. The success of generative models for data augmentation may depend on the specific task.

To achieve a compromise between data diversity and implementation difficulty, methods based on combining existing annotated data have been developed for data augmentation, which are easy to implement and allow more variability of the generated data than data augmentation based on basic image transformation. For example, Mixup linearly combines two annotated images and the corresponding annotations~\cite{Mixup}, and the combination is performed stochastically to create a large number of augmented training images. CutMix is further developed to allow nonlinear combination of two images, where one region in the combined image is from one image and the rest is from the other image~\cite{Cutmix}. The annotations are still linearly combined according to the contribution of each image~\cite{Cutmix}. 
However, these data augmentation approaches based on image combination are mostly applied to image classification problems~\cite{bdair2020roam,2020Analyzing}, and the development of this type of data augmentation methods for brain lesion segmentation is still an open problem.

In this work, we develop a data augmentation approach that produces diverse training data and is easy to implement for brain lesion segmentation.
Similar to Mixup and CutMix, the proposed method combines existing annotated data for the generation of new training data; unlike these methods, the combination in our method is lesion-aware, and thus the proposed method is more appropriate for brain lesion segmentation. Specifically, given a pair of annotated training images, from one image we carve a \textit{region of interest}~(ROI) according to the lesion location and geometry, and then the carved region replaces the corresponding voxels in the other labeled image. 
The size of the ROI is sampled from a probability distribution so that diverse combinations can be achieved.
The annotation of the second image in this region is replaced by the corresponding labels in the first image as well. Since the combination is achieved with a carving operation, our method is referred to as CarveMix.
To evaluate our method, experiments were performed for two brain lesion segmentation tasks, where CarveMix was integrated with the state-of-the-art segmentation framework nnU-Net~\cite{0nnU} and improved the segmentation accuracy. 
The codes of our method are available at \url{https://github.com/ZhangxinruBIT/CarveMix.git}.

\section{Method}
\subsection{Problem Formulation}

Suppose we are given a set $\mathcal{X}=\{\mathbf{X}_{i}\}_{i=1}^{N}$ of 3D annotated images with brain lesions, where $\mathbf{X}_{i}$ is the $i$-th image and $N$ is the total number of images. The annotation of $\mathbf{X}_{i}$ is denoted by $\mathbf{Y}_{i}$, and the set of annotations is denoted by $\mathcal{Y}=\{\mathbf{Y}_{i}\}_{i=1}^{N}$.
In this work, we consider binary brain lesion segmentation, and thus the intensity of $\mathbf{Y}_{i}$ is either 1 (lesion) or 0 (background).

$\mathcal{X}$ and $\mathcal{Y}$ can be used to train a CNN that automatically segments brain lesions. 
In addition, it is possible to perform data augmentation to generate new images and annotations from $\mathcal{X}$ and $\mathcal{Y}$, so that more training data can be used to improve the network training.
It is shown in classification problems and some segmentation problems that the combination of pairs of existing annotated images is a data augmentation approach that can generate diverse training data and is also easy to implement~\cite{Mixup,Cutmix}.
Specifically, from an image pair $\mathbf{X}_{i}$ and $\mathbf{X}_{j}$ as well as the pair of annotations $\mathbf{Y}_{i}$ and $\mathbf{Y}_{j}$, a synthetic image $\mathbf{X}$ and its annotation $\mathbf{Y}$ are generated.
By repeating the image generation for different image pairs and different sampling of generation parameters, a number of synthetic images and annotations can be created and used together with $\mathcal{X}$ and $\mathcal{Y}$ to improve network training.
However, existing methods based on image combination are not necessarily appropriate for brain lesion segmentation problems, because their design is unaware of lesions and the generation of annotations is designed for classification problems.
Therefore, it is desirable to develop an effective combination-based data augmentation approach for brain lesion segmentation. 

\subsection{CarveMix}

To develop a data augmentation approach based on image combination for brain lesion segmentation, we propose CarveMix, which is lesion-aware and thus more appropriate for brain lesion segmentation. 
In CarveMix, given $\{\mathbf{X}_{i},\mathbf{Y}_{i}\}$ and $\{\mathbf{X}_{j},\mathbf{Y}_{j}\}$, we carve a 3D ROI from $\mathbf{X}_{i}$ according to the lesion location and geometry and mix it with $\mathbf{X}_{j}$. 
Specifically, the extracted ROI replaces the corresponding region in $\mathbf{X}_{j}$, and the replacement is also performed for the annotation using $\mathbf{Y}_{i}$ and $\mathbf{Y}_{j}$.
Mathematically, the synthetic image $\mathbf{X}$ and annotation $\mathbf{Y}$ are generated as follows
\begin{eqnarray}
\label{eq:carve1}
\mathbf{X} &=& \mathbf{X}_{i}\odot \mathbf{M}_{i} + \mathbf{X}_{j}\odot (1 - \mathbf{M}_{i}), \\
\mathbf{Y} &=& \mathbf{Y}_{i}\odot \mathbf{M}_{i} + \mathbf{Y}_{j}\odot (1 - \mathbf{M}_{i}).
\label{eq:carve2}
\end{eqnarray}
Here, $\mathbf{M}_{i}$ is the ROI (binary mask) of extraction determined by the lesion location and geometry given by the annotation $\mathbf{Y}_{i}$, and $\odot$ denotes voxelwise multiplication. 

As mentioned above, the ROI $\mathbf{M}_{i}$ should be lesion-aware. Thus, $\mathbf{M}_{i}$ is designed to extract the ROI along the lesion contours. In addition, to allow more diversity of the extracted ROI and thus the generated image, the size of $\mathbf{M}_{i}$ is designed to be randomly sampled from a probability distribution.
To this end, we first compute the signed distance function $D(\mathbf{Y}_{i})$ for the lesion regions of $\mathbf{X}_{i}$ using its annotation $\mathbf{Y}_{i}$, where the intensity $D^{v}(\mathbf{Y}_{i})$ of $D(\mathbf{Y}_{i})$ at voxel $v$ is determined as 
\begin{equation}
D^{v}(\mathbf{Y}_{i})=\left\{  
\begin{aligned}
-d(v,\partial{\mathbf{Y}_{i}}),\ \mathrm{if} \  \mathbf{Y}_{i}^{v} = 1  &  \\ 
 d(v,\partial{\mathbf{Y}_{i}}), \ \mathrm{if} \ \mathbf{Y}_{i}^{v} = 0  &
\end{aligned}
\right. .
\label{eq:sdf}
\end{equation}
Here, $\partial{\mathbf{Y}_{i}}$ represents the boundary of the lesions in $\mathbf{Y}_{i}$, $d(v,\partial{\mathbf{Y}_{i}})$ represents the distance between $v$ and the lesion boundary, and $\mathbf{Y}_{i}^{v}$ denotes the intensity of $\mathbf{Y}_{i}$ at $v$. Then, we can obtain $\mathbf{M}_{i}$ that is consistent with the location and shape of the lesions by thresholding $D(\mathbf{Y}_{i})$, where the value $\mathbf{M}^{v}_{i}$ of $\mathbf{M}_{i}$ at voxel $v$ is
\begin{equation}
\mathbf{M}^{v}_{i}=\left\{  
\begin{aligned}
1,&\ D^{v}(\mathbf{Y}_{i}) \leq \lambda   \\ 
0,& \ \mbox{otherwise}
\end{aligned}
\right. .
\label{eq:threshold}
\end{equation}
Here, $\lambda$ is a threshold that is sampled from a predetermined distribution.
A greater $\lambda$ leads to a larger carved ROI.
To allow the carved region to be larger or smaller than the lesion, $\lambda$ can be either positive or negative, respectively.
Thus, the distribution for sampling $\lambda$ is defined as a mixture of two uniform distributions
\begin{equation}
\lambda \sim \frac{1}{2}U(\lambda_{\mathrm{l}},0) + \frac{1}{2}U(0,\lambda_{\mathrm{u}}),
\label{eq:lambda}
\end{equation}
where $\lambda_{\mathrm{l}}$ and $\lambda_{\mathrm{u}}$ are the lower and upper bounds of the distribution $U(\lambda_{\mathrm{l}},0)$ and $U(0,\lambda_{\mathrm{u}})$, respectively.

Since the minimum value $D(\mathbf{Y}_{i})_{\mathrm{min}}$ of $D(\mathbf{Y}_{i})$ is an indicator of the lesion size, $\lambda_{\mathrm{l}}$ and $\lambda_{\mathrm{u}}$ are determined adaptively based on $D(\mathbf{Y}_{i})_{\mathrm{min}}$ as 
\begin{equation}
\lambda_{\mathrm{l}}=-\frac{1}{2}|D(\mathbf{Y}_{i})_{\mathrm{min}}| \mbox{ and } \lambda_{\mathrm{u}}=|D(\mathbf{Y}_{i})_{\mathrm{min}}|.
\end{equation}
In this way, the relative variation of the ROI size with respect to the lesion size is within two.
Then, we have
\begin{equation}
\lambda \sim \frac{1}{2}U(-\frac{1}{2}|D(\mathbf{Y}_{i})_{\mathrm{min}}|,0) + \frac{1}{2}U(0,|D(\mathbf{Y}_{i})_{\mathrm{min}}|).
\label{eq:sample}
\end{equation}
\begin{figure}[!t]
\includegraphics[width=\textwidth]{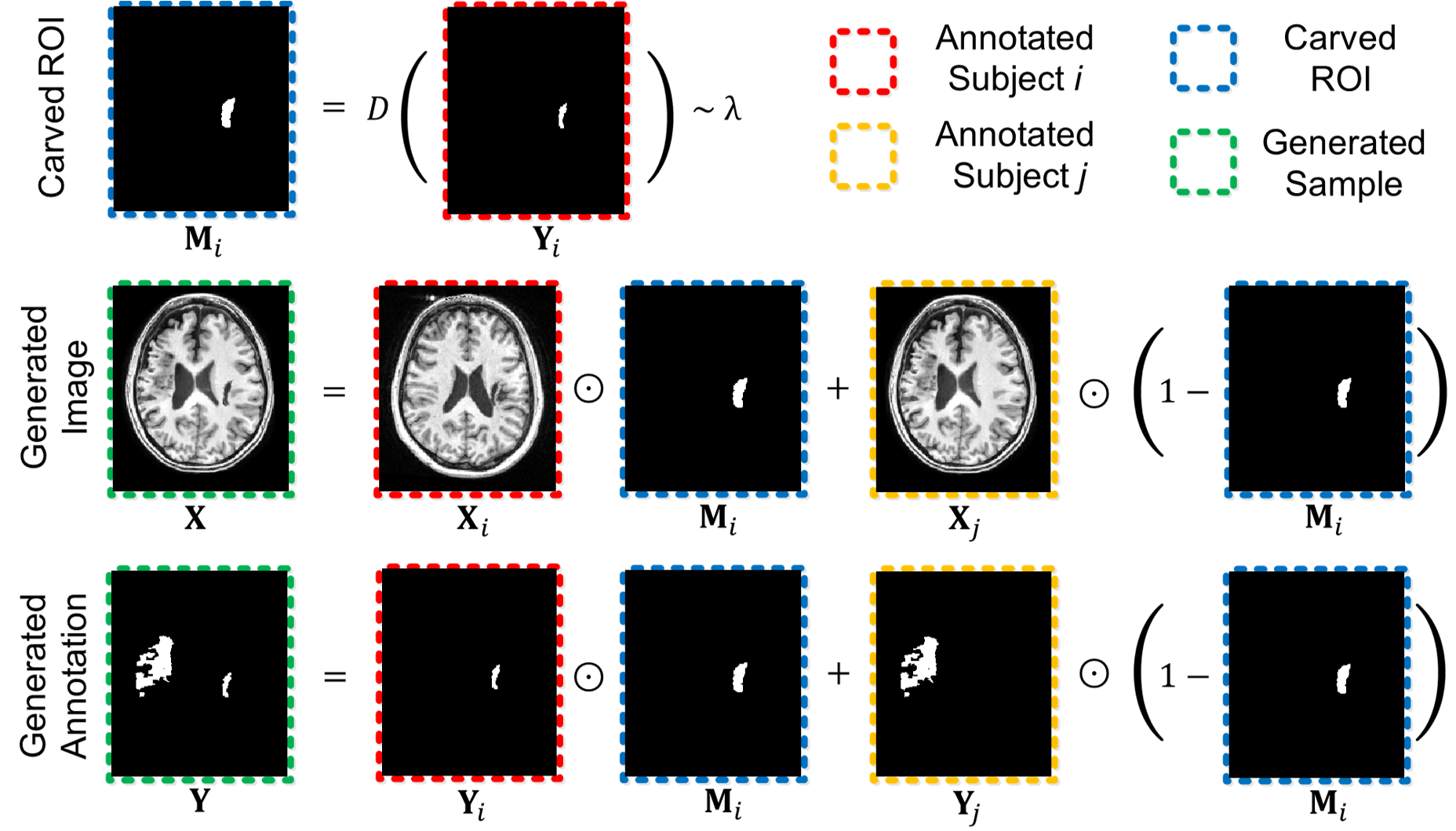}
\caption{A graphical illustration of the image generation procedure in CarveMix.} \label{fig:graphical_illustration}
\end{figure}

A graphical illustration of the CarveMix procedure described above is shown in Fig.~\ref{fig:graphical_illustration}, where a synthetic image and its annotation are generated from a pair of annotated images and their annotations. 
This procedure can be repeated by randomly drawing pairs of annotated images and the annotations as well as the size parameter $\lambda$ for each image pair, so that a set $\mathcal{X}_{\mathrm{s}}$ of synthetic images and the corresponding set $\mathcal{Y}_{\mathrm{s}}$ of annotations can be generated for network training. 
The complete CarveMix algorithm for generating the sets of synthetic images and annotations is summarized in Algorithm~\ref{algo:CarveMix}.
Note that like in Mixup and CutMix, the generated images may not always look realistic.
However, existing works have shown that unrealistic synthetic images are also able to improve network training despite the distribution shift, and there exists a tradeoff between the distribution shift and augmentation diversity~\cite{gontijo2021tradeoffs}.
\floatname{algorithm}{Algorithm}
\renewcommand{\algorithmicrequire}{\textbf{Input:}}
\renewcommand{\algorithmicensure}{\textbf{Output:}}
\begin{algorithm}[!t]
\caption{CarveMix}
\label{algo:CarveMix}
\begin{algorithmic} 
	\Require Training images $\mathcal{X}$ and annotations $\mathcal{Y}$; the desired number $T$ of synthetic images
	\Ensure Synthetic images $\mathcal{X}_{\mathrm{s}}$ and annotations $\mathcal{Y}_{\mathrm{s}}$
	\For{ $t = 1,2,...,T$ }:   
	
	Randomly select a pair of training subjects:  $\{\mathbf{X}_{i},\mathbf{Y}_{i}\}$ and $\{\mathbf{X}_{j},\mathbf{Y}_{j}\}$ 
	
    Compute the signed distance function $D(\mathbf{Y}_{i})$ for $\mathbf{Y}_{i}$ with Eq.~(\ref{eq:sdf})
    
	Sample $\lambda$ using Eq.~(\ref{eq:sample})
    
    Threshold $D(\mathbf{Y}_{i})$ with $\lambda$ to obtain a carved ROI $\mathbf{M}_{i}$ according to Eq.~(\ref{eq:threshold})
	
	Generate a synthetic image and its annotation using Eqs.~(\ref{eq:carve1}) and (\ref{eq:carve2})
	
	\EndFor\State
	\Return $\mathcal{X}_{\mathrm{s}}$ and $\mathcal{Y}_{\mathrm{s}}$
	
\end{algorithmic}
\end{algorithm}

\subsection{Relationship with Mixup and CutMix}
\label{sec:relation}

\begin{table}[!t]

\centering
\caption{Comparison of Mixup, CutMix, and CarveMix}
\label{tab:summary}
\resizebox{\textwidth}{!}{
\begin{tabular}{cccc}  
\toprule
Method    & Image Generation     & Annotation Generation & Notes\\  
\toprule        
Mixup     & $\mathbf{X}=\lambda\mathbf{X}_{i} + (1- \lambda)\mathbf{X}_{j} $          
          & $\mathbf{Y}=\lambda\mathbf{Y}_{i} + (1- \lambda)\mathbf{Y}_{j} $
          &\makecell[c]{$\lambda$ is sampled from \\the beta distribution} \\
\hline
CutMix    & $\mathbf{X} = \mathbf{X}_{i}\odot \mathbf{M}_{i} + \mathbf{X}_{j}\odot(1 -  \mathbf{M}_{i})$
          & $\mathbf{Y}= \lambda\mathbf{Y}_{i} + (1- \lambda)\mathbf{Y}_{j} $
          &\makecell[c]{ $\mathbf{M}_{i}$ is a randomly \\ selected  cube and \\  $\lambda$ is determined by \\ the size of $\mathbf{M}_{i}$}
          \\
\hline
CarveMix  & $\mathbf{X} = \mathbf{X}_{i}\odot \mathbf{M}_{i} + \mathbf{X}_{j}\odot(1 - \mathbf{M}_{i})$                 
          &\quad$\mathbf{Y} = \mathbf{Y}_{i}\odot \mathbf{M}_{i} + \mathbf{Y}_{j}\odot   (1 - \mathbf{M}_{i})$ 
          &\makecell[c]{$\mathbf{M}_{i}$ is selected \\ according to the lesion \\ location and shape}\\

\bottomrule
\end{tabular}
}
\end{table}

The setup in Eqs.~(\ref{eq:carve1}) and (\ref{eq:carve2}) bears similarity with the Mixup~\cite{Mixup} and CutMix~\cite{Cutmix} frameworks, where synthetic samples are also generated from pairs of annotated training subjects. However, Mixup and CutMix may not be suitable for brain lesion segmentation.
To see that, we summarize and compare the data generation procedures in Mixup, CutMix, and CarveMix in Table~\ref{tab:summary}. Note that for Mixup and CutMix the generation of annotations is extended to voxelwise combination for 3D image segmentation.
Both Mixup and CutMix are unaware of lesions, where the data generation does not pay special attention to the lesions. 
In addition, in CutMix the synthetic intensity at each voxel originates from one individual image, but the generation of labels simply linearly combines the annotations of the two images at the voxel instead of using the label of the subject that contributes to the voxel. 
CarveMix addresses these limitations for brain lesion segmentation, where the generation of images is lesion-aware and the generation of annotations is consistent with the image generation.

\subsection{Implementation Details}

The proposed method can be used for either online or offline data augmentation.
In this work, we choose to perform offline data augmentation, where a desired number $T$ of synthetic samples are generated before network training, and these samples are combined with true annotated data to train the segmentation network. In this way, our method is agnostic to the segmentation approach.
For demonstration, we integrate CarveMix with the state-of-the-art nnU-Net method, which has achieved consistent top performance for a variety of medical image segmentation tasks~\cite{0nnU} with carefully designed preprocessing and postprocessing.\footnote{Note that CarveMix can also be integrated with other segmentation frameworks if they are shown superior to nnU-net.}
nnU-Net uses the U-net architecture~\cite{2dunet,3dunet} and automatically determines the data configuration, including intensity normalization, the selection of 2D or 3D processing, the patch size and batch size, etc.
For more details about nnU-Net, the readers are referred to~\cite{0nnU}.
The default hyperparameters of nnU-Net are used, except for the number of training epochs, because we empirically found that a smaller number was sufficient for training convergence in our experiments (see Sect.~\ref{sec:eval}).

\section{Experiments}

\subsection{Data Description}

To evaluate the proposed method, we performed experiments on two brain lesion datasets, where chronic and acute ischemic stroke lesions were segmented, respectively. 
The first dataset is the publicly available ATLAS dataset~\cite{2018A} for chronic stroke lesions, which contains 220 annotated T1-weighted images. These images have the same voxel size of 1 mm isotropic. 
We selected 50 images as the test set and considered several cases for the training set, where different numbers of the remaining images were included in the training set.
Specifically, in these cases 170, 85, 43, and 22 annotated training images were used, which corresponded to 100\%, 50\%, 25\%, and 12.5\% of the total number of the available annotated images, respectively.
For each case, 20\% of the images in the training set were further split into a validation set for model selection.

The second dataset is an in-house dataset for acute ischemic stroke lesions, which includes 219 annotated \textit{diffusion weighted images}~(DWIs). The DWIs were acquired on a 3T Siemens Verio scanner with a $b$-value of 1000 s/$\mbox{mm}^2$.
The image resolution is $0.96 \mbox{ mm} \times 0.96 \mbox{ mm}  \times 6.5 \mbox{ mm}$.
We selected 50 images as the test data and considered four cases of the training set, where 169 (100\%), 84 (50\%), 42 (25\%), and 21 (12.5\%) annotated training images were used, respectively. For each case, 20\% of the training images were further split into a validation set.

\subsection{Evaluation Results}
\label{sec:eval}

\begin{figure}[!t]
\includegraphics[width=\textwidth]{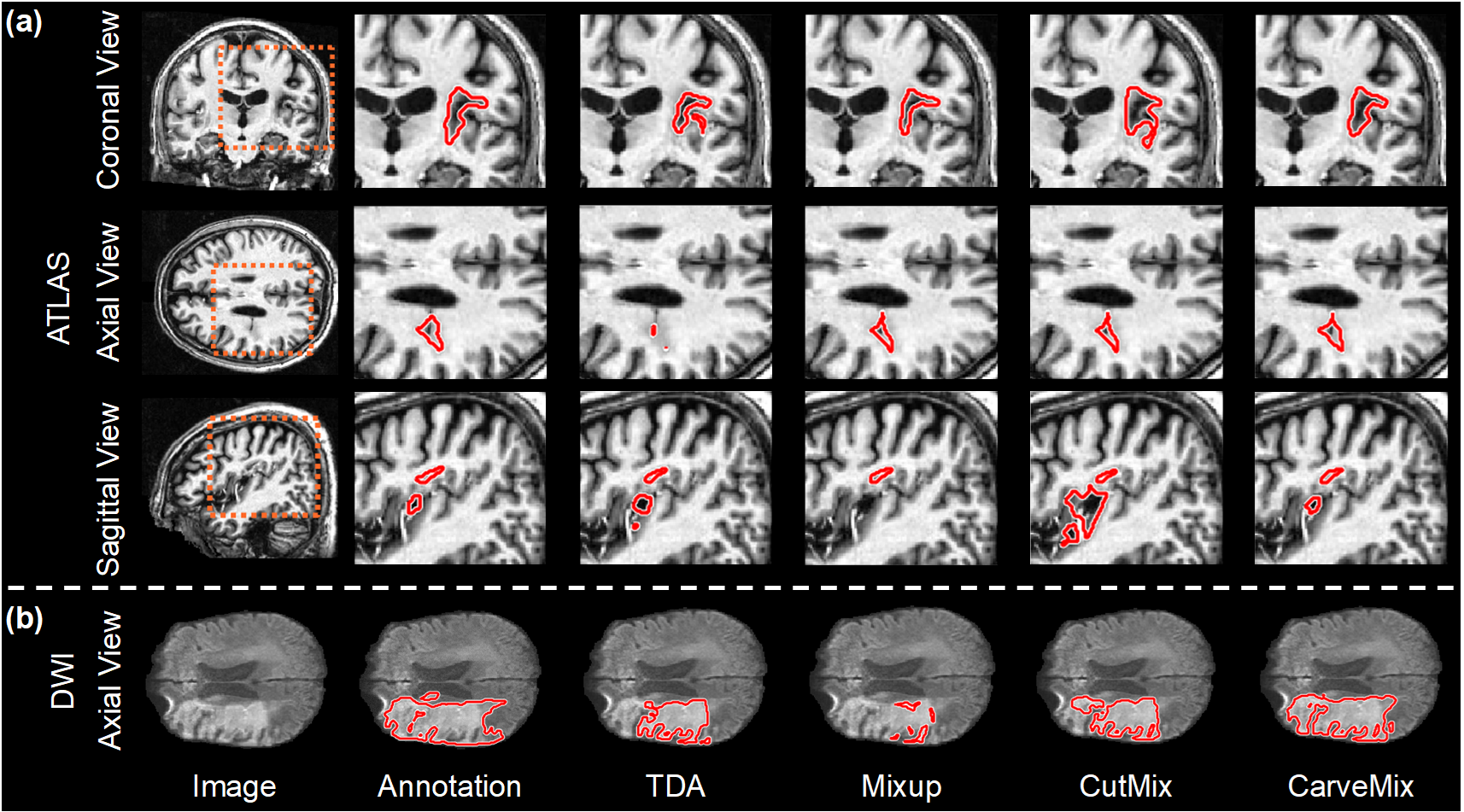}
\caption{Cross-sectional views of the segmentation results (red contours) on representative test scans: (a) triplanar views and their zoomed views for the ATLAS dataset and (b) axial views for the DWI dataset. The results of CarveMix and each competing method are overlaid on the T1-weighted image or DWI for the ATLAS or DWI dataset, respectively, and they are shown together with the expert annotation.
These results were obtained with 100\% annotated training scans.}
\label{ATLAS_DWI}
\end{figure}

CarveMix was applied to the two datasets separately for each experimental setting.
It was compared with the default \textit{traditional data augmentation}~(TDA) implemented in nnU-Net~\cite{0nnU}, including rotation, scaling, mirroring, elastic deformation, intensity perturbation, and simulation of low resolution. CarveMix was also compared with Mixup~\cite{Mixup} and CutMix~\cite{Cutmix} with their default hyperparameters.
For CarveMix, Mixup, and CutMix, the synthetic annotated scans were generated so that a total number of 1000 scans (including the true annotated scans) were available for training.
CarveMix, Mixup, and CutMix were integrated with nnU-Net offline, and thus TDA was also performed for these synthetic images.
Note that since the same number of epochs and the same number of batches per epoch were used for each method during network training, and TDA was performed randomly online, all methods including TDA have used the same number of training samples. Therefore, the comparison with TDA was fair.
The maximum number of training epochs was set to 450/200 for the ATLAS/DWI dataset, respectively.
The evaluation results are presented below.

\subsubsection{Results on the ATLAS dataset}

We first qualitatively evaluated CarveMix. Triplanar views of the segmentation results on a representative test scan are shown in Fig.~\ref{ATLAS_DWI}(a) for CarveMix and each competing method, together with the expert annotation.
Here, the results were obtained with 100\% training data (170 annotated training scans). 
We can see that CarveMix produced segmentation results that better agree with the annotation than the competing methods.

\begin{table}[t]
\centering
\caption{Means and standard deviations of the Dice coefficients (\%) of the segmentation results on the test set for the ATLAS/DWI dataset. The results for each size of the training set are shown. Asterisks indicate that the difference between the proposed method and the competing method is statistically significant (*: $p \le 0.05$, **: $p\le 0.01$, ***: $ p\le 0.001$) using a paired Student's $t$-test. The best results are highlighted in bold.}
\label{ATLAS_DWI_result}
\resizebox{\textwidth}{!}{
\begin{tabular}{ccllll}  
\toprule
Dataset &
\multicolumn{1}{c}{\quad Size}&
\multicolumn{1}{c}{TDA}&
\multicolumn{1}{c}{Mixup}&
\multicolumn{1}{c}{CutMix}&
\multicolumn{1}{c}{\quad CarveMix}\\
\midrule        
\multirow{4}*{ATLAS}
~&\quad100\%  &\quad $59.39 \pm 32.45 ^*$          &\quad$59.33 \pm33.06 ^*$ 
              &\quad $56.11 \pm 32.44 ^{**}$       &\quad \bm{$63.91 \pm 29.87$} \\
~&\quad50\%   &\quad $56.72 \pm 30.74 $            &\quad$58.40 \pm29.35 $   
              &\quad $54.25 \pm 30.24 ^*$          &\quad \bm{$60.57 \pm 31.77$} \\
~&\quad25\%   &\quad $49.87 \pm 32.19 ^{***}$      &\quad$49.18 \pm32.72 ^{***}$ 
              &\quad $41.19 \pm 33.98 ^{***}$      &\quad \bm{$55.82 \pm 31.58$} \\
~&\quad12.5\% &\quad $41.86 \pm 32.87^{***}$       &\quad$42.57 \pm33.54 ^{***}$ 
              &\quad $24.57 \pm 27.01 ^{***}$      &\quad \bm{$54.77 \pm 30.55$} \\
\hline
\multirow{4}*{DWI}
~&\quad100\%  &\quad$74.91 \pm25.22 ^*$    &\quad$74.19 \pm25.22 $
              &\quad$73.33 \pm27.30 ^{*}$  &\quad \bm{$76.40 \pm25.31$} \\
~&\quad50\%   &\quad$73.35 \pm25.91 ^*$    &\quad$71.10 \pm27.50^* $
              &\quad$69.70 \pm27.36 ^{**}$ &\quad \bm{$74.99 \pm25.34$} \\
~&\quad25\%   &\quad$69.41 \pm27.94 ^*$    &\quad$68.71 \pm28.56 ^* $
              &\quad$50.28 \pm32.44 ^{***}$&\quad \bm{$72.07 \pm26.64$}\\
~&\quad12.5\% &\quad$64.83 \pm25.23 ^{**}$ &\quad$57.04 \pm31.86 ^{***}$
              &\quad$07.72  \pm15.31 ^{***}$&\quad \bm{$71.32 \pm24.59$} \\

\bottomrule
\end{tabular}
}
\end{table}

Next, CarveMix was quantitatively evaluated. For each method and each experimental setting of the training set, we computed the means and standard deviations of the Dice coefficients of the segmentation results on the test set. These results are summarized in Table~\ref{ATLAS_DWI_result} (the part associated with the ATLAS dataset). 
In all cases, CarveMix outperforms the competing methods with higher Dice coefficients.
In addition, in most cases the difference between CarveMix and the competing methods is statistically significant using paired Student's $t$-tests, and this is also indicated in Table~\ref{ATLAS_DWI_result}. 
Note that Mixup and CutMix are not originally designed for brain lesion segmentation. Compared with TDA they do not necessarily improve the segmentation quality, which is consistent with the previous observations in~\cite{2020Semi}. The CutMix strategy could even degrade the segmentation performance due to the inappropriate generation of synthetic annotations discussed in Sect.~\ref{sec:relation}. 

\subsubsection{Results on the DWI dataset}

Similar to the evaluation on the ATLAS dataset, qualitative and quantitative evaluation was performed for the DWI dataset. The results are shown in Fig.~\ref{ATLAS_DWI}(b) and Table~\ref{ATLAS_DWI_result} (the part associated with the DWI dataset). 
From Fig.~\ref{ATLAS_DWI}(b) we can see that the result of CarveMix better resembles the expert annotation than those of the competing methods; and Table~\ref{ATLAS_DWI_result} indicates that CarveMix has better Dice coefficients than the competing methods and its difference with the competing methods is significant in most cases.

\section{Conclusion}
We have proposed CarveMix, which is a simple data augmentation approach for brain lesion segmentation. 
The proposed method combines pairs of annotated training samples to generate synthetic training images, and the combination is lesion-aware.
The experimental results on two brain lesion segmentation tasks show that CarveMix improves the segmentation accuracy and compares favorably with competing data augmentation strategies.

\subsubsection{Acknowledgements}
This work is supported by Beijing Natural Science Foundation (L192058 \& 7192108).

\bibliographystyle{splncs04}
\bibliography{refs}

\end{document}